\documentclass[prl,twocolumn,amsfonts,showpacs,a4paper]{revtex4}

\usepackage{amsmath}
\usepackage{graphicx}

\def\be{\begin{equation}}
\def\ee{\end{equation}}
\def \bea#1\eea {\begin{eqnarray}#1\end{eqnarray}}

\def\br{{\bf r}}

\def\ETh{E_\text{Th}}
\def\Re{\mathop{\rm Re}}

\begin{document}

\title{Proximity-induced superconductivity in graphene}

\author{M. V. Feigel'man}
\email{feigel@landau.ac.ru}
\author{M. A. Skvortsov}
\author{K. S. Tikhonov}
\affiliation{L. D. Landau Institute for Theoretical Physics, Moscow 119334, Russia}
\affiliation{Moscow Institute of Physics and Technology, Moscow 141700, Russia}

\date{October 1, 2008}

\begin{abstract}
We propose a way of making graphene superconductive by putting on it
small superconductive islands which cover a tiny fraction of
graphene area. We show that the critical temperature, $T_c$,
can reach several Kelvins
at the experimentally accessible range of parameters.
At low temperatures, $T \ll T_c$, and zero magnetic field,
the density of states is characterized by a small gap $E_g \leq T_c$
resulting from the collective proximity effect.
Transverse magnetic field $H_{g}(T) \propto E_g$ is expected to destroy
the spectral gap driving graphene layer to a
kind of a superconductive glass state.
Melting of the glass state into a metal occurs
at a higher field $H_{g2}(T)$.
\end{abstract}



\maketitle

Among numerous fascinating properties,
 graphene~\cite{1,2} provides a unique possibility to study the phenomenon of
 proximity-induced superconductivity in very favorable conditions.
 Experimental studies of the Josephson current through graphene
 in standard wide planar SNS junctions~\cite{graphene-Jos}
 have shown that proximity effect in graphene is
 qualitatively similar to the one known for usual dirty metals.
In this Letter, we show that even a tiny amount of graphene area covered by
 small superconductive islands  (with good electric contact to graphene)
 can lead to a macroscopically superconductive state of the graphene film,
 with $T_c$ in the Kelvin range.

\begin{figure}
\includegraphics[width=0.95\columnwidth]{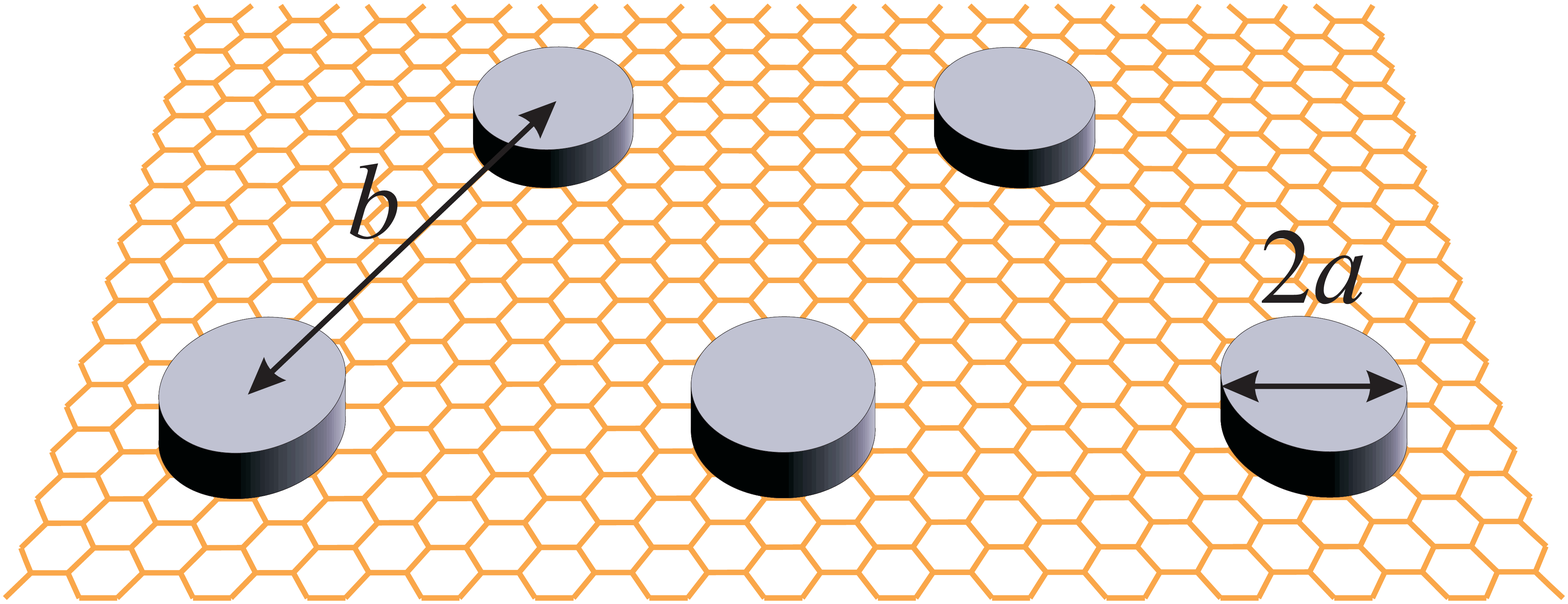}
\caption{\label{F:setup}(Color online) Graphene film covered by superconducting islands.}
\end{figure}

We consider a system of superconductive (SC) islands of radius $a$
(with the typical value of few tenths of nanometer)
placed approximately uniformly on top of a graphene
 layer (with the typical
 distance between islands $b$ in the sub-micron range)
shown in  Fig.~\ref{F:setup}.
We assume that $b$ is much
 larger than both $a$ and the graphene mean-free-path $l$.
 Moreover, present theory will be limited by the case
 $l \lesssim a$ when electron motion in graphene is
 diffusive at all relevant scales.
 We will not be particulary interested in phenomena in the vicinity of
 the graphene
 neutral point,
 assuming relatively large gate potentials $|V_g| \geq 10$ V, and carrier
 density $n \geq 10^{12}$ cm$^2$. We assume  graphene Fermi energy
$E_F \gg \Delta_0 \gg T_c$,
 where $\Delta_0$ is the island's superconductive gap.
Graphene sheet can be either single- or few-layered:
 the  only relevant features
 are (i) high diffusion constant $D \geq 10^2$ cm$^2$/s, and (ii) very low
 (in comparison with metals)
 electron density, which allows to combine
 moderate values of dimensionless conductance
 $g =  (\hbar/e^2R_\Box) \geq 3$ with high Thouless energy
$E_{\rm Th} = \hbar D/b^2$.
Not very large values of sheet conductance $g$
are practically favorable to avoid suppression
of superconductivity in small SC islands due to
the inverse proximity effect \cite{comment-inv-prox}.

Below we treat graphene as a normal diffusive  2D metal within the standard approach
based on the Usadel equation~\cite{Usadel}; its applicability to diffusive graphene
was proven in Ref.~\cite{Titov,Tikhonov}. The intrinsic Cooper channel interaction
in graphene can be neglected due to its low DOS \cite{comment-intrinsic-Cooper}.
Similarly, phonon-induced attraction is also weak.

{\em Proximity coupling and transition temperature.}---%
We start with calculating the Josephson
coupling energy between two superconductive
islands of radius $a$ separated by distance $b \gg a$,
neglecting the presence of other islands.
Such a pair-wise approximation is adequate for determination of $T_c$,
but breaks down at $T \leq T_c/\ln(b/a)$, as shown below.
For the single SC island on graphene, the
Matsubara-space Usadel equation  for the spectral angle
$\theta_\omega$ and  corresponding boundary conditions~\cite{KuprLuk}
read as
\begin{gather}
\label{Usadel}
  D\nabla^2\theta_\omega - 2|\omega|\sin\theta_\omega = 0 ,
\\
\left. \left[ g\frac{\partial\theta_\omega}{\partial r}
+\frac{G_{\rm int}}{2\pi a}\cos\theta_\omega\right]\right|_{r=a} =0
\label{KuLu} .
\end{gather}
The normal ($G_\omega$) and anomalous ($F_\omega$) components
of the matrix Green function the in Nambu-Gorkov space
are expressed via the spectral angle $\theta_\omega$ and the
order parameter phase $\varphi$ as
$G_\omega({\bf r}) = \cos\theta_\omega({\bf r})$ and
$F_\omega({\bf r}) = e^{i\varphi}\sin\theta_\omega({\bf r})$.
The full matrix structure of the anomalous Green function $\check{F}_\omega$
with the valley and spin spaces included
(Pauli matrices $\hat{\pi}$ and $\hat{s}$, respectively)
is determined by the usual $s$-wave pairing in the SC islands:
$\check{F}_\omega \propto \hat{\pi}_x \hat{s}_y$.
The interface conductance $G_{\rm int}$
is treated below as a phenomenological parameter which accounts for
Fermi velocity mismatch and a potential barrier on graphene-metal
interface~\cite{GG08}.

It is crucial for further analysis that the two-island generalization
of the nonlinear problem (\ref{Usadel}), (\ref{KuLu}) can be linearized
while calculating the Josephson current at inter-island distances $b \gg a$.
Indeed, the total current can be calculated by integrating the current density
over the middle line between the islands, on the distance $\rho_{1,2} \geq b/2$
from them. This procedure also involves summation over Matsubara energies
$\omega_n = \pi T(2n+1)$, with the major contribution to the sum
coming from $\omega_n \sim E_{\rm Th}$. At such $\omega_n$ and $\rho_{1,2}$
the spectral angle $\theta$ is small, and linearization of Eqs.~(\ref{Usadel})
and (\ref{KuLu}) leads to the solution
\begin{equation}
\theta_\omega (r) = A(\omega)K_0\left(\frac{r}{L_\omega}\right) ,
\qquad
A(\omega) = \frac{\Theta(t_\omega)}{\ln(L_\omega/a)} ,
\label{theta-sol}
\end{equation}
with $L_\omega = \sqrt{D/2\omega}$,
$t_\omega = (G_{\rm int}/2\pi g)\ln(L_\omega/a)$,
and $\Theta(t)$ solving the equation $\Theta(t) = t\cos\Theta(t)$.
The function $A(\omega)$ evolves between the tunnel and diffusive
limits as
\be
  A(\omega)
  =
  \begin{cases}
    G_{\rm int}/(2\pi g), & G_{\rm int}\ll 2\pi g/\ln(L_\omega/d) , \\
    \pi/[2\ln(L_\omega/d)], & G_{\rm int}\gg 2\pi g/\ln(L_\omega/d) ,
  \end{cases}
\label{A}
\ee
and is always small for $\ln(L_\omega/a) \gg 1$.
Thus the Josephson current $I(\varphi)=I_c\sin\varphi$ between two SC islands
with different phases, $\varphi_1-\varphi_2=\varphi$,
can be calculated using the linearized two-island solution for the anomalous
Green function:
$
F_\omega({\bf r}) = e^{i\varphi_1}\sin\theta_\omega(|{\bf r}-{\bf r}_1|)
+ e^{i\varphi_2}\sin\theta_\omega(|{\bf r}-{\bf r}_2|)
$.
The standard calculation of the
Josephson energy $E_J = (\hbar/2e)I_c$ then leads to
\be
\label{EJ(r)}
  E_J(b,T)
  =
  4\pi g T\sum_{\omega_n > 0} A^2(\omega_n) \,
  P\bigl(\sqrt{\omega_n/8\ETh}\bigr) ,
\ee
where
$P(z) = z \int_0^\infty K_0(z\cosh t) K_1(z\cosh t) \, dt$.

\begin{figure}
\includegraphics[width=0.95\columnwidth]{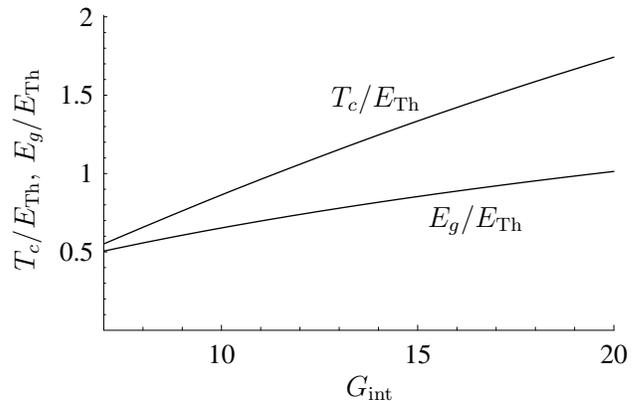}
\caption{The critical temperature, $T_c$, and the zero-temperature spectral gap,
$E_g,$ vs.\ the interface conductance $G_{\rm int}$
(the sheet conductance $g=6$, and $b/a = 10$).}
\label{TcvsG}
\end{figure}

A two-dimensional array of SC islands with the coupling energies (\ref{EJ(r)})
undergoes the Berezinsky-Kosterlitz-Thouless transition at
\be
\label{BKT}
  T_c = \gamma \, E_J(b,T_c) ,
\ee
where the numerical coefficient $\gamma$ depends on the array structure.
Below we will assume that the SC islands
form a triangular lattice, in which case $\gamma \approx 1.47$
\cite{Stroud}.
For the interface conductance $G_{\rm int}$ comparable with
the sheet conductance $g$, one finds the transition temperature
$T_c \sim \ETh$.
In general, $T_c$ can be obtained by numerical solution of
Eq.~(\ref{BKT}) using Eqs.~(\ref{theta-sol}) and (\ref{EJ(r)}).
The result obtained for the ratio  $T_c/E_{\rm Th}$ as a function
of $G_{\rm int}$   for $g=6$ ($R_\Box\approx700\,\Omega$)
and $b/a = 10$ is presented in Fig.~\ref{TcvsG}.
With the graphene diffusion constant $D = 500$ cm$^2$/s
(see, e.g., \cite{Savchenko2007})
and $b=0.5$ $\mu$m, one estimates
$E_{\rm Th} \approx 1.5 K$, leading to $T_c$ in the range $1 \div 3$ K
for $5<G_{\rm int}<20$.

{\em Low temperatures: spectral gap and order parameter.}
Now we switch to the low-temperature range $T \ll T_c$  and consider
the issue of the spectral gap for the excitation above
the fully coherent ground state (with all phases $\varphi_i$ equal).
The density of states $\nu(E) = \nu_0\Re\cos\theta(E)$ is determined
then by the periodic solution of Eqs.~(\ref{Usadel}) and (\ref{KuLu}),
analytically continued to real energies: $|\omega|\to iE$.
This periodic problem is equivalent to the one defined within
the single (hexagonal) elementary cell, supplemented by the
additional condition
$\mathbf{n\nabla}\theta\bigr|_{\br\in\Gamma} = 0$,
where $\Gamma$ is the cell boundary.
Solution of the Usadel equations for such a geometry leads to
formation of the spectral gap $E_g$ similar
to the minigap for one-dimensional SNS junctions~\cite{SNS}.
To find it, we write $\theta(\br) = \pi/2 + i\psi(\br)$
and determine the spectral boundary as the value of $E$ where equation
\be
D\nabla^2\psi+2E\cosh\psi=0
\label{Usadel-psi}
\ee
ceases to have solutions with real $\psi({\bf r})$ \cite{OSF2001}.
At large $\ln(b/a)$ one may approximate the hexagonal boundary $\Gamma$
of the elementary cell by the circle of radius $R = b/2$.
For the ideally transparent interface, numerical solution
of the radially symmetric equation (\ref{Usadel-psi})
gives for the value of the zero-temperature spectral gap:
\be
  E_g \approx \frac{\hbar D/R^2}{1.52\ln(R/a)- 1.2}
  \approx
  \frac{2.6 E_{\rm Th}}{\ln(b/4a)} .
\label{Eg}
\ee
Decreasing the interface conductance $G_{\rm int}$ leads to the suppression
of the the minigap, as shown in Fig.~\ref{TcvsG}.

In the limit of large $\ln(b/a)$, the spectral gap $E_g \ll T_c$.
Smallness of the gap distinguishes
the system with superconductive islands from usual dirty superconductors.
Roughly speaking, it behaves as a continuous 2D superconductor
at the energy/temperature scales smaller than $E_g$,
whereas in the range $E_g < (E,T) < T_c$ it can rather be
described as an array of weak Josephson junctions.

The existence of the sharp gap (\ref{Eg}) in the electron spectrum
looks surprising, as only a tiny fraction $(a/b)^2$ of graphene
area is in direct contact with SC islands.
The presence of this gap can be traced back
to the periodic structure of islands we assumed.
Therefore any irregularity in the positions of SC islands
will lead to the smearing of the hard gap.
Assuming that islands' locations are shifted at random from the sites of
the ideal triangular lattice, with the typical shift $\delta b \ll b$,
one can reduce the problem to
the effective one, defined on a scales large than array lattice constant.
Random displacements of islands will be seen, in terms of this
effective model, as local fluctuations of the superconductive
coupling constant~\cite{LO72,MeyerSimons01},
leading to the smearing of the gap with the relative
width $\delta E_g \sim (\delta b/b)^2 E_g$.
The sharp gap will also be smeared by thermal fluctuations of
island's phases $\varphi_i$ and  finite thermal coherence length
$L_T$.  Thus we expect the spectral gap to be observable at $T \ll E_g$.

Even in the presence of the gap smearing, strong suppression of the local
DoS in graphene at $E \leq E_g$, should be seen by the low-temperature Scanning
Tunnelling Microscopy. The spectral (pseudo) gap is a signature of collective
proximity effect
which cannot be quantitatively described by a pair-wise interaction
between SC islands as soon as low-energy scales $\leq E_g$ are involved.
The corresponding spatial scale
\be
  \xi_g
  =
  \sqrt{\hbar D/E_g}
  \sim
  b \sqrt{\ln (b/a)}
\ee
plays the role of low-temperature coherence length in the
(dirty-limit)
superconductor.
Under our main condition $\ln(b/a) \gg 1$, the coherence length
$\xi_g \gg b$,
which allows continuous treatment of the array at low temperatures.

The local superconductive order parameter in graphene,
${\cal F}({\bf r}) = \int d\omega F_\omega({\bf r})$,
can be found at $T < E_g$ as
\be
{\cal F}({\bf r}) =
  \sum_j  \frac{D e^{i\varphi_j}}{({\bf r}-{\bf r}_j)^2}
  \frac{\Theta[t(|{\bf r}-{\bf r}_j|)]}{\ln(|{\bf r}-{\bf r}_j|/a)} ,
\label{Fr}
\ee
where
${\bf r}_j$ are the coordinates of SC islands,
$t(r)= (G_{\rm int}/2\pi g)\ln(r/a)$,
and we used the solution (\ref{theta-sol})
for $\omega > E_g$.
The divergent sum in Eq.~(\ref{Fr}) should be cut at $|\br-\br_j|\sim\xi_g$
since the spectral gap (\ref{Eg}) suppresses the lowest-$\omega$ contribution
to ${\cal F}({\bf r})$.
Equation~(\ref{Fr}) is not applicable in the vicinity of SC islands
since Eqs.~(\ref{Usadel}) and (\ref{KuLu})
cannot be linearized at small $|{\bf r}-{\bf r}_j|$.

At zero magnetic field, $\varphi_j= {\rm const}$
and the space-averaged order parameter $\overline{\cal F}$
is given by
\be
\overline{\cal F} = n_{\rm i}
\int_b^{\xi_g} d^2r\frac{D}{r^2}\frac{\Theta[t(r)]}{\ln(r/a)}
= \frac{\pi^2}{2} n_{\rm i}D\frac{\ln\ln(b/a)}{\ln(b/a)} ,
\label{Fav}
\ee
where the last expression
refers to the large-$G_{\rm int}$ limit ($\Theta \approx \pi/2$)
and $n_{\rm i} \approx 1/b^2$ is the concentration of SC islands.
Comparison of (\ref{Fav}) and (\ref{Eg}) provides  the condition
for neglecting the intrinsic Cooper-channel interaction in graphene, $\lambda_g$.
Namely, its presence would generate the the energy
gap $\Delta_g = \lambda_g\overline{\cal F}$. This ``intrinsic'' gap
can be neglected compared to proximity-induced gap (\ref{Eg}) provided that
$\lambda_g \ll 0.5$. Comparison with the estimate for intrinsic Cooper
interaction constant \cite{comment-intrinsic-Cooper}
shows that the latter is indeed negligible.

{\em Electromagnetic response.}---%
Linear response of the superconductive film to a weak electromagnetic field is
characterized by the superconductive density $\rho_s$.
In the intermediate temperature range, $E_g \ll T \ll T_c$,
one can easily calculate $\rho_s$ within the
pair-wise approximation for the proximity coupling:
\be
  \rho_s(T)
  =
  \frac{n_i^2}{2} \int_0^\infty 2\pi r^3 E_J(r,T)\, dr .
\label{rhos}
\ee
Taking $E_J(r,T)$ from Eq.~(\ref{EJ(r)}) we find
\be
  \rho_s(T) = (\pi^3/3) g A^2(\pi T) \ETh^2/T ,
\label{rhos1}
\ee
where $g=2\nu_0D$ and the numerical factor corresponds
to the triangular array with $n_{\rm i} = (2/\sqrt{3})b^{-2}$.
At lowest temperatures, $T < E_g$, the function $\rho_s(T)$
saturates at the value that can be
estimated by the replacement $T \to E_g$ in Eq.~(\ref{rhos1}).
For highly transparent interface we obtain
\be
  \rho_s(0)
  \approx
  10 g \ETh /\ln(b/a) .
\label{rhos2}
\ee
Comparing (\ref{rhos2}) and (\ref{Eg}) we find that
$\rho_s(0) \approx 4 g E_g$, which is
typical for dirty superconductors with the gap $E_g$.

The critical current density per unit length
at the lowest temperatures,
$T \ll E_g$, can be estimated as
\be
  j_c(0)
  \approx
  \frac{2e}{\hbar b}E_J(b,0)
  =
  \frac{\pi^3}{2}\frac{e g D}{b^3\ln^2(b/a)} ,
\label{jc}
\ee
where we used, as an estimate, the $T=0$ limit of the pair-wise
Josephson coupling energy (\ref{EJ(r)}) at $G_{\rm int}/g \gg 1$:
\be
  E_J(b,0) = \frac{\pi^3}{4}\frac{ g E_{\rm Th}}{\ln^2(b/a)} .
\label{EJ0}
\ee

\begin{figure}
\includegraphics[width=\columnwidth]{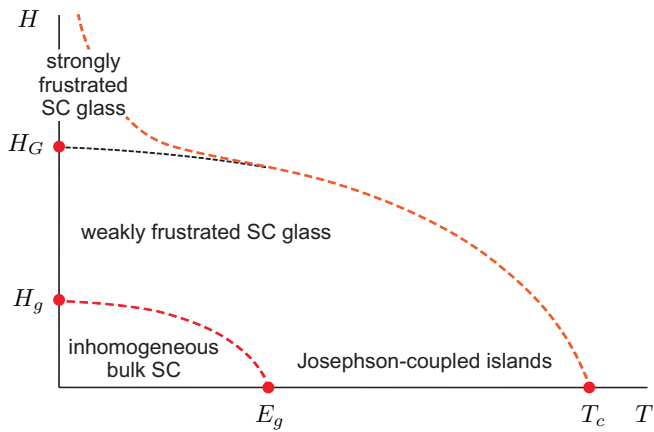}
\caption{Schematic phase diagram of the graphene sheet with superconductive islands; all
lines refer to crossovers rather than to sharp phase transitions.}
\label{F:phase-diag}
\end{figure}

{\em The effect of the transverse magnetic field}
is characterized by two different field scales:
\be
H_{g} = \frac{\Phi_0}{2\pi\xi_g^2} \approx
\frac{0.4}{\ln(b/4a)}\frac{\Phi_0}{b^2} ,
\qquad
H_{\rm Glass} = \frac{\Phi_0}{b^2} .
\label{critfields}
\ee
In the low-field, low-$T$ region ($B \ll H_{g}$ and  $T \ll E_g$),
magnetic field produces well-separated
pancake ``hyper-vortices'' with the core size $\xi_g \geq b$
(the local DoS is gapless in the core regions).
These vortices are strongly pinned by the underlying array structure,
so a high critical current
$j_{c1} \sim j_c(0) (b/\xi_g)$ is expected.
 At $B \approx H_{g}$ vortex cores overlap and the
proximity gap is totally destroyed, so $H_g$ is an analogue of the upper
critical field $H_{c2}$.
However, the metallic state is not formed right above $H_g$,
at  least at $T \ll E_g$.
In this field range one deals with a system of frustrated pair-wise
Josephson couplings, with full frustration achieved at $B \gg H_{\rm Glass}$.
In this high-field range, average values of Josephson coupling
are exponentially suppressed,
$\overline{E_J(B)} \propto \exp(-B /H_{\rm Glass})$.
However, as shown in \cite{SpivakZhou,GalLar01}, actual (random-sign)
Josephson couplings are much stronger due to mesoscopic fluctuations:
\be
  E^{\rm glass}_J(b)
  =
  \bigl[\overline{(E_J(b))^2}\bigr]^{1/2}
  \sim
  \frac{E_{\rm Th}}{\ln^2(b/a)} ,
\label{EJglass}
\ee
which is just by the factor $1/g$ smaller than the $(T,B)=0$ pairwise
 coupling (\ref{EJ0}).
The estimate (\ref{EJglass}) shows that at $T \to 0$
the superconductive glass state
survives up to high magnetic fields $H_{g2}(0) \gg H_{\rm Glass}$.
 The value of $H_{g2}(0)$ is determined
by quantum phase fluctuations~\cite{GalLar01,FLS2001}:
\be
  H_{g2} \sim (\Phi_0/b^2) e^{c\sqrt{g}} , \quad c \sim 1 \, .
\label{Hg2}
\ee
The overall phase diagram in the $(H,T)$ plane is shown
schematically in Fig.~\ref{F:phase-diag}. We emphasize that the
lines indicated do not refer to sharp phase transitions
(which are absent in the presence of magnetic field, apart from
some special values of frustration for the case of well-defined lattice of islands)
but rather mark a crossover regions. Note that determination of $T_c$
for a rational values of frustration $f=\frac14,\frac13,\frac12$
(where results for nearest-neighbor XY model are available) is complicated by
the necessity of accounting for long-range proximity couplings.

To conclude,
we have shown that graphene can be made superconductive with $T_c$ of
the order of few Kelvins, due to collective
proximity effect induced by small superconductive islands covering
only a tiny part of graphene sheet area.
The spectral gap is expected at low temperatures $T \ll T_c$
and magnetic fields $B \ll \Phi_0/b^2$.
Transformation from a continuous disordered superconductive state to a
weakly coupled junction array is predicted with the temperature
and/or magnetic field increase.

Our study was based on the standard Usadel equations
which are valid for sufficiently disordered samples ($l \ll a$).
In the opposite limit of quasi-ballistic electron motion around islands,
the Andreev subgap conductance might
decrease, leading to the increase of quantum fluctuations of phases
$\varphi_i(t)$ \cite{FLS2001,FL1998}.
Quantum fluctuations can be neglected under the condition
$b^2\ln(b/a) \ll b_c^2$, where $b_c$ is the critical distance between
the islands marking the quantum phase transition (QPT) to the metallic
state~\cite{FLS2001}.
 The same problem of suppressed Andreev conductance
 appears to be even more serious with the decrease of the electron
density towards the graphene neutral point: scattering cross-section
of electrons on the SC islands drops in the range of $k_F a \sim 1$
 leading to effective decoupling of island's phases
and to strong quantum fluctuations.  These fluctuations may lead to
the QPT of the superconductor--metal type~\cite{FLS2001,FL1998}.

We are grateful to P. A. Ioselevich, P. M. Ostrovsky and M. Titov
for useful discussions.
This work was partially supported by RFBR Grant No.\ 07-02-00310.


\begin{thebibliography}{99}
\bibitem{comment-inv-prox}
The inverse proximity effect can be neglected provided that
$G_{\rm tot}\delta \leq \Delta_0$, where
$G_{\rm tot}^{-1} = G^{-1}_{\rm int} + \ln(b/a)/2\pi g  \ll 1$
is the total normal-state resistance from the island to the
graphene sheet, $G_{\rm int}$ is the SC-graphene interface conductance,
and $\delta$ is level spacing in the island.
\bibitem{comment-intrinsic-Cooper}
In graphene, repulsion in the Cooper channel involves electrons
from different valleys (wavevectors near the $K$ and $K'$ points),
which means that the Coulomb scattering amplitude is $U_C= 2\pi e^2/(\kappa K_0)$,
where $\kappa \approx 5$ is the effective dielectric constant
of graphene on the substrate and $K_0= 1.7\cdot 10^8$ cm$^{-1} $ is the
distance between $K$ and $K'$ points in the reciprocal space.
The dimensionless (repulsive) coupling constant is very small,
$-\lambda_g = U_C \nu_0 = (2e^2/\kappa\hbar v_F)(k_F/K_0) \leq 0.03$
for $n = k_F^2/\pi < 10^{13}$ cm$^{-2}$,
due to the DOS $\nu_0 = k_F/\pi\hbar v_F$
per single spin projection.
\end{thebibliography}
\end{document}